\documentstyle[11pt,aaspp4,psfig]{article}

\def\rxte{{\it RXTE}}

\begin{document}

\lefthead{Miller}
\righthead{Brightness Oscillations From 4U~1636--536}

\title{A Characterization of the Brightness Oscillations 
 During Thermonuclear Bursts From 4U 1636--536}

\author{M.\ Coleman Miller}
\affil{Department of Astronomy and Astrophysics, University of Chicago\\
       5640 South Ellis Avenue, Chicago, IL 60637, USA\\
       miller@bayes.uchicago.edu}
\authoremail{miller@bayes.uchicago.edu}

\begin{abstract}

The discovery of nearly coherent brightness oscillations
during thermonuclear X-ray bursts from six neutron-star
low-mass X-ray binaries has opened up a new way to study the
propagation of thermonuclear burning, and
may ultimately lead to greater
understanding of thermonuclear propagation in other
astrophysical contexts, such as in Type Ia supernovae.
Here we report
detailed analyses of the $\sim$580~Hz brightness oscillations
during bursts from 4U~1636--536.  We investigate the
bursts as a whole and, in more detail, the initial portions
of the bursts.  We analyze the $\sim$580~Hz
oscillations in the initial 0.75 seconds of the five bursts 
that were used in a previous search for a brightness oscillation
at the expected $\sim$290~Hz spin frequency,
and find that if the same frequency model describes all five
bursts there is insufficient data to require more than a
constant frequency or, possibly, a frequency plus a frequency
derivative.  Therefore, although it is appropriate to use
an arbitrarily complicated model of the $\sim$580~Hz
oscillations to generate a candidate waveform for the
$\sim$290~Hz oscillations, models with more than two
parameters are not required by the data.  For the bursts as a
whole we show that the characteristics of the brightness
oscillations vary greatly from burst to burst. We find,
however, that in at least one of the bursts, and possibly in
three of the four that have strong brightness oscillations
throughout the burst, the oscillation frequency reaches a
maximum several seconds into the burst and then decreases.
This behavior has not been reported previously for burst
brightness oscillations, and it poses a challenge to the
standard burning layer expansion explanation for the frequency
changes.

\end{abstract}


\section{INTRODUCTION}

Shortly after the launch of the Rossi X-ray Timing Explorer (\rxte)
in December 1995, observation with \rxte\ of neutron-star
low-mass X-ray binaries (LMXBs) revealed that several sources had a 
single, highly coherent, high-amplitude
brightness oscillation during at least one thermonuclear X-ray burst (for
reviews see Strohmayer, Zhang, \& Swank 1997; Strohmayer, Swank,
\& Zhang 1998a).  The asymptotic frequency
of these oscillations in the tails of bursts is so similar in
different bursts from a single source, and the oscillation
is so coherent in the tail (see, e.g., Strohmayer \& Markwardt
1999), that it is almost certain that this
asymptotic frequency is the stellar spin frequency or its first
overtone.  These burst oscillations therefore provided the first
direct evidence for the value of the spin frequencies of these
LMXBs, and they corroborate strongly the proposed evolutionary link
between LMXBs and millisecond rotation-powered pulsars.  In addition,
the stability of the frequency in the tails of the bursts has
led to their application as promising probes of the binary systems
themselves (Strohmayer et al.\ 1998b).

The existence of burst oscillations indicates that the
emission from the surface, and hence the thermonuclear
burning, is not uniform over the entire star.  This is in
accord with theoretical expectations (Joss 1978; Ruderman 1981;
Shara 1982; Livio \& Bath 1982; Fryxell \& Woosley 1982;
Nozakura, Ikeuchi, \& Fujimoto 1984; Bildsten 1995), and it suggests
that the properties of the burst oscillations, such as the
evolution of their
frequency or amplitude, may contain valuable information
about the propagation of thermonuclear burning over the
surface of the neutron star.  The lessons learned
from study of the thermonuclear propagation in bursts may
ultimately further our understanding of thermonuclear propagation
in other astrophysical contexts, such as classical novae
and Type Ia supernovae.  Unlike in
novae or Type Ia supernovae, burning in thermonuclear X-ray
bursts occurs near the surface and occurs often for a single
source, and is therefore relatively easy to observe.  The detailed
study of burst brightness oscillations therefore has broad
importance.

Here we describe in detail the frequency behavior of the burst
oscillations in five bursts from 4U~1636--536, which is an
LMXB with an orbital period of 3.8 hours
(see, e.g., van Paradijs et al.\ 1990).  This source is of
special interest because it produces detectable signals at
both the fundamental and the first overtone of the stellar
spin frequency (Miller 1999), and because near the beginning
of one burst the brightness oscillations reached the highest
amplitude ---50\% rms--- so far recorded for oscillations 
during a thermonuclear burst (Strohmayer et al.\ 1998c).

In \S~2 we analyze the light curves of the bursts, 
and the frequency and amplitude of the brightness oscillations
in the four of those five bursts that have strong brightness
oscillations for most of the duration of the burst.  
We find that, despite apparent
similarities in the light curves of three of those four bursts, the
amplitude and frequency behavior of their brightness oscillations
are very different from each other.  We also find compelling evidence
in one burst, and strong evidence in another burst, for an
interval in which the burst oscillation frequency decreases after
the peak in the light curve.  

In \S~3 we focus on the initial portions of the bursts.  Analyses
of bursts from many sources have shown that the oscillation
frequency often changes by a few Hertz over the first few
seconds of a burst (see Strohmayer et al.\ 1998a for a review).  
The change is often a monotonic rise,
but there are indications of more complicated behavior in some
bursts.  It has been pointed out that the magnitude of the
frequency change could be explained by a 20--50 meter expansion
of the burning layers followed by a slow settling, if the layers
conserve angular momentum (see, e.g., Strohmayer et al.\ 1998a),
but details have not been worked out.  For example, it is not clear
how the layers would maintain their coherence throughout the 5--10
complete circuits relative to the body of the star that are implied by
the observations.  Bildsten (1998) has suggested that the layers
may be stabilized by thermal buoyancy or mean molecular weight
stratification, but details have not been worked out.
In \S~3 we examine in detail the first
0.75 seconds of all five bursts, which was the interval used
to construct the candidate waveform for the $\sim$290~Hz
oscillation in 4U~1636--536 (Miller 1999).  We examine models
of the frequency behavior that have increasing complexity:
a constant-frequency model; one with a frequency and frequency
derivative; a four-parameter model with an 
initial frequency and frequency derivative
followed by a different frequency derivative after a break time;
and a five-parameter model with two different frequencies and
frequency derivatives separated by a break time.  We find that
if the same type of frequency model applies to all five of
the bursts then the data do not require a model more complicated
than the constant-frequency model or, possibly, the model
with a single frequency and frequency derivative.  Note, however,
that this is not inconsistent with the use of the five-parameter
model to construct a waveform used in the search for the
expected $\sim$290~Hz oscillation (Miller 1999); in such a search,
the only goal is to find the best fit to the $\sim$580~Hz
oscillations, and the extra parameters need not be justified
by a significantly better fit.

Finally, in \S~4 we discuss the implications of 
these results for the current picture of the frequency changes,
in which the frequency change occurs because the burning layer
is lifted by 20--50 meters from the surface by the radiation
flux.  We find that the simplest version of this picture has
difficulty explaining the observations.

\section{OVERVIEW OF THE BURSTS}

\begin{figure}
 \psfig{file=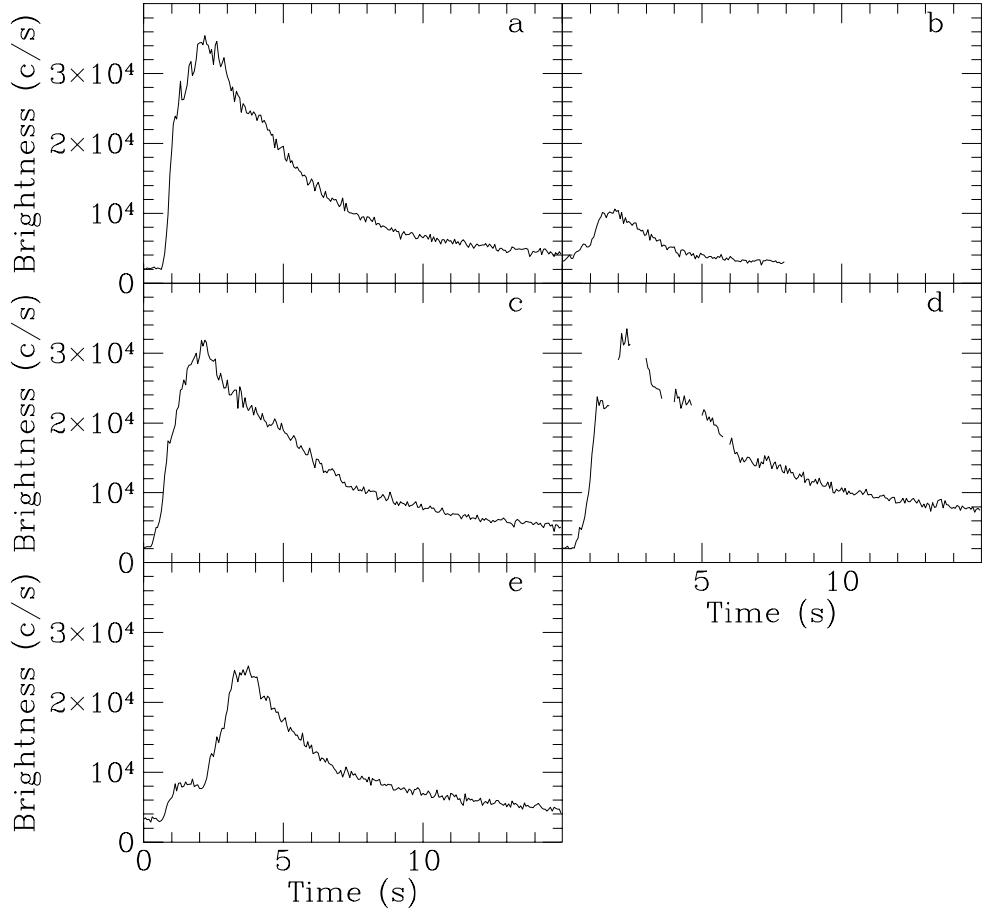,height=6.0truein,width=6.0truein}
 \caption[]{
 \label{figLightcurves}
Light curves for the bursts.  (a)~Burst beginning at
22:39:24 UTC on 28 December 1996. (b)~Burst beginning at
23:54:02 UTC on 28 December 1996. (c)~Burst beginning at
23:26:46 UTC on 29 December 1996. (d)~Burst beginning at
17:36:52 UTC on 31 December 1996. (e)~Burst beginning at
09:57:26 UTC on 23 February 1998. The data gaps in burst~d
are caused by telemetry saturation.}
 \end{figure}

We used public-domain data
from the High Energy Astrophysics Science Archive Research
Center.  The data were taken in Single Bit Mode, which does
not record the energy of photons.  We give
the starting times of the bursts in Table~1 and the
light curves in Figure~\ref{figLightcurves}.
In burst~d, the data dropouts are caused by telemetry saturation.

\begin{table*}[t] 
\centering
\begin{tabular}{cc}
\multicolumn{2}{c}{\bfseries Table~1: 
Starting Time of Bursts}{\rule[-2mm]{0mm}{6mm}}\\
\tableline
\tableline
Burst&Date and Time\\
\tableline
a&22:39:24.188 UTC on 28 December 1996\\
b&23:54:02.876 UTC on 28 December 1996\\
c&23:26:46.813 UTC on 29 December 1996\\
d&17:36:52.941 UTC on 31 December 1996\\
e&09:57:25.938 UTC on 23 February 1997\\
\tableline

\end{tabular}
\end{table*}

\begin{figure}
 \psfig{file=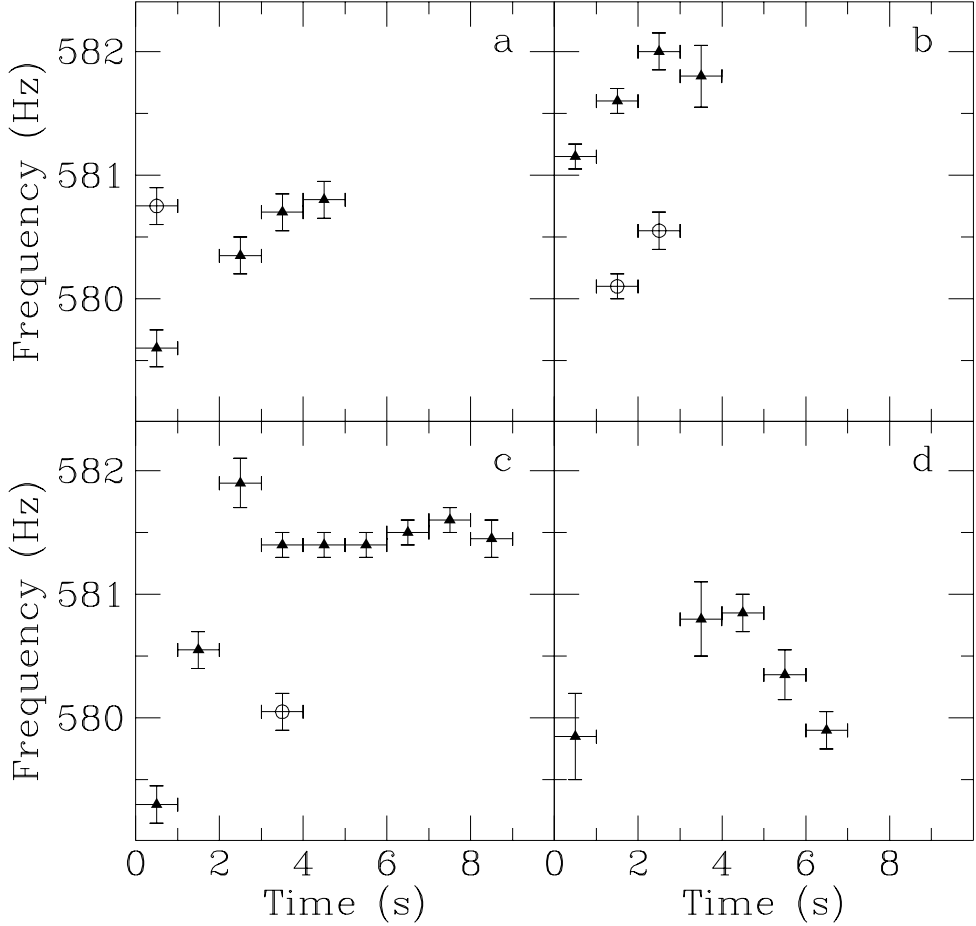,height=6.0truein,width=6.0truein}
 \caption[]{
 \label{figDynamic}
Power spectra as a function of time for the four bursts with strong
brightness oscillations. Each solid triangle
is at the frequency
of maximum power and its $1\sigma$ uncertainties for nonoverlapping
1~second intervals.  The frequency of a peak is only plotted
if its Leahy power exceeds 10.  In bursts 1, 2, and 3 there are
intervals in which a second peak exceeds this threshold, and this
secondary peak is plotted with an open circle.  The power of
the secondary peak in burst~(a) is 21.2 (single-trial
significance $2.5\times 10^{-5}$); the powers of
the secondary peaks in burst~(b) are 44.0 (significance
$2.8\times 10^{-10}$) and 13.5 (significance $1.2\times 10^{-3}$); and
the power of the peak in burst~(c) is 12.8 (significance
$1.7\times 10^{-3}$).  The panels are labeled as in
Figure~\ref{figLightcurves}.  Burst~(e) does not have
strong brightness oscillations for most of the burst, and is
therefore excluded from this analysis.}
 \end{figure}

\begin{figure}
 \psfig{file=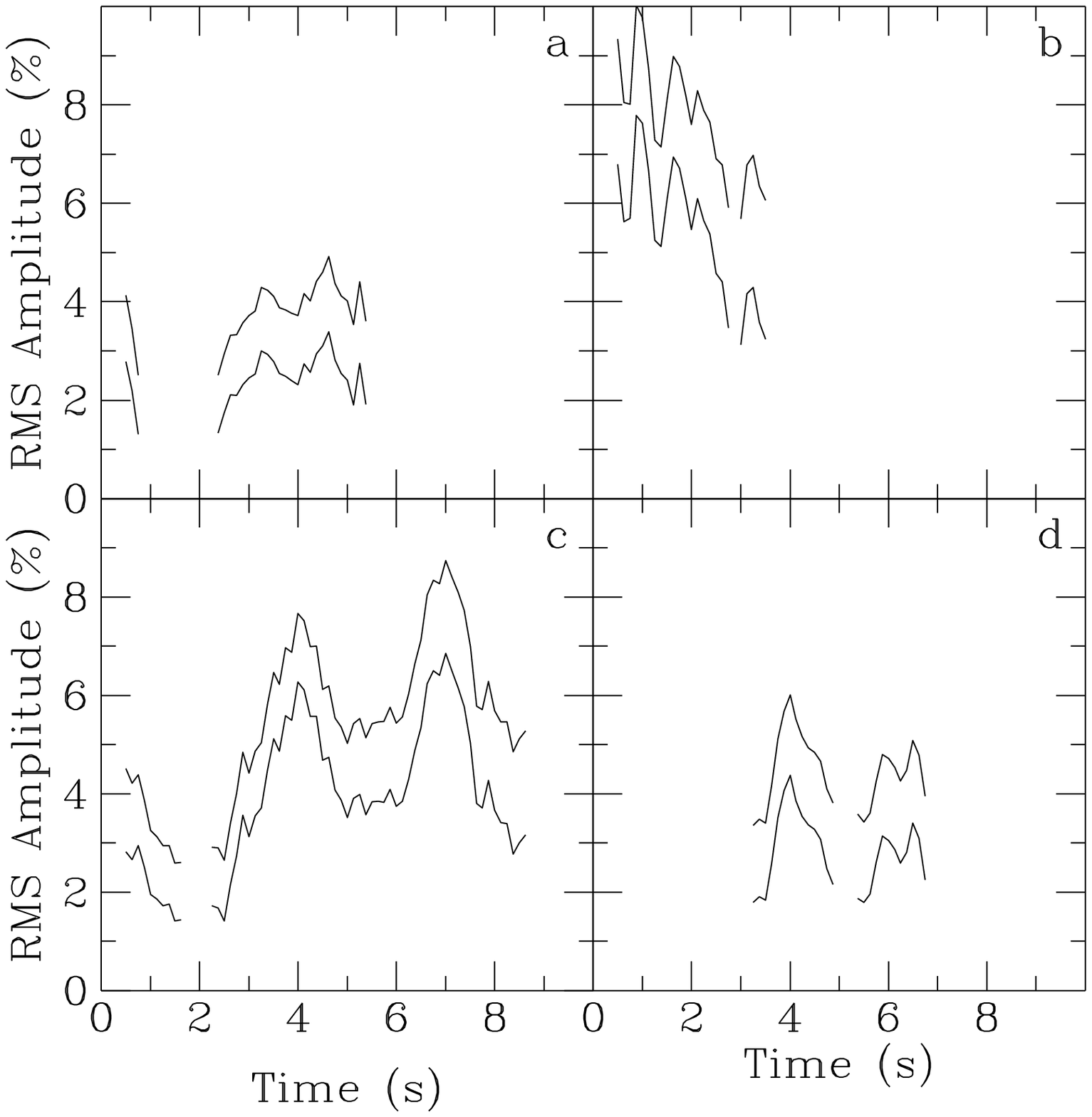,height=6.0truein,width=6.0truein}
 \caption[]{
 \label{figAmplitudes}
Root mean square amplitude of the brightness oscillation
for each of the bursts.  The amplitudes are calculated for
one-second intervals with starting times 1/8 second apart,
and the $\pm 1\sigma$ uncertainty bands are shown.  The 
amplitude is only plotted if
the Leahy power for the oscillation exceeds 10.  The panels
are labeled as in Figure~\ref{figLightcurves}.}
 \end{figure}

Figure~\ref{figDynamic} shows the peaks of the power spectra
of the first four bursts, as a function of time.  The burst on 
23 February 1997 does not have a strong brightness oscillation 
for most of its duration, and we therefore do not analyze it in the
rest of this section.  For each burst, the frequency of maximum
power in successive nonoverlapping one-second intervals is
shown by the solid triangles, and the Leahy et al.\ (1983)-normalized
power at this frequency is shown by the solid line.  Here we plot
only those points with Leahy powers in excess
of 10 (chance probability for a single trial less than
$6\times 10^{-3}$).  The horizontal bars on the frequency points
indicate the extent of the interval for which the power
density spectrum was calculated.  In a few cases, 
more than one peak exceeds this
threshold in a given power density spectrum.  We then
represent the lower-power peak by an open circle.  In burst~a
the secondary peak has a Leahy power of 21.2 (single-trial
significance $2.5\times 10^{-5}$); in burst~b the
secondary peaks have Leahy powers of 44.0 (first interval;
significance $2.8\times 10^{-10}$)
and 13.5 (second interval; significance $1.2\times 10^{-3}$); 
and in burst~c the secondary peak has a Leahy power of 12.8
(significance $1.7\times 10^{-3}$).
Finally, Figure~\ref{figAmplitudes} shows the rms amplitude of
each oscillation, computed for one-second intervals 1/8 second apart.

It is evident from these figures that the frequency behavior
can be very complex and can differ greatly from burst to burst.
The light curves for bursts~(a), (c), and (d)
appear similar to each other, although burst~(d) has a
slightly longer decay time than the other two.  However,
the frequency and amplitude of the brightness oscillations 
evolve very differently in the three bursts.  

In burst~(a),
there is a strong oscillation near the beginning
which disappears for approximately one second, then the
oscillation reappears after the peak.  The frequency 
increases continuously, although there is some evidence
that in the initial $\sim$0.5 second of the burst the
frequency drops (this might help explain the presence of a
higher-frequency secondary peak in the power density spectrum).

In burst~(c), the brightness oscillation is present for
almost the entire time examined.  The frequency increases rapidly
in the first two to three seconds, then appears to decrease
to an asymptotic value.  A power density spectrum of a two-second interval 
starting 1.75 seconds after the beginning of the burst reveals
a peak at 581.62$\pm$0.04~Hz.  A power density spectrum of a
six-second interval starting four seconds after the beginning
of the burst has a peak at 581.47$\pm$0.01~Hz.  If the latter
frequency is the asymptotic frequency of the oscillation, then
at a 3$\sigma$ level of certainty it is less than the maximum frequency
attained during the burst.  The amplitude of the oscillation
in the burst tail is high and significant, and there is an
abrupt increase in the amplitude 6--8 seconds after the beginning
of the burst that is not accompanied by any apparent change
in the light curve.

In burst~(d) there is a clear {\it decrease} in the
frequency of the burst oscillation in the tail of the burst.
We explored this further by taking a power density spectrum of
a longer interval: five seconds, starting three seconds after
the beginning of the burst.  We found that, at the 99.99\%
confidence level, the frequency change per second during this
interval is $-0.54\pm 0.08$~Hz~s$^{-1}$.  The best-fit 
frequency at the beginning of this five-second interval 
depends on the frequency derivative, and is approximately
$\nu_0=581.39{\rm\ Hz}-2({\dot\nu}+0.62~{\rm Hz~s}^{-1})
{\rm\ Hz}$.  This means that, relative to a brightness oscillation
with a constant frequency equal to the frequency at the
beginning of this five-second interval, the observed brightness
oscillation has a total phase lag of between $12\pi$ and
$16\pi$ radians.  The total phase lag is comparable to what
is seen in many bursts, except that here the frequency inferred
in the tail of the burst is significantly less than the
spin frequency inferred from other bursts in this source.
There is no sign in this burst that the frequency has reached
an asymptotic value.

Burst~(b) is the only one with a clearly different
light curve.  This is a weak burst.  The frequency of the
brightness oscillation is consistent with what is observed
in, at least, bursts~(a) and (c): a rise in the
frequency near the beginning of the burst, followed by an
approximate leveling off.  We note, however, that within
the uncertainties the frequency could also reach a maximum
and then decline, as appears to be the case for bursts~(c)
and (d).

\section{BRIGHTNESS OSCILLATIONS AT THE BEGINNING OF THE BURST}

Previous analyses have shown that the brightness
oscillations in the initial
$\sim$second of the bursts are often of particular interest.
This is where the highest amplitudes (rms$\sim$50\%; Strohmayer
et al.\ 1998c) are reported, and where subharmonics of the
strong oscillation have been detected in 4U~1636--536
(Miller 1999) and possibly in the Rapid Burster (Fox \& Lewin
1999).  It is therefore important to examine the initial portion
more closely to see what hints about the brightness oscillation
mechanism can be derived.

Before doing so, we need to emphasize an important distinction.
If the purpose of the analysis is to
characterize the frequency variations of the $\sim$580~Hz
oscillation then extra parameters can only be added if the
fit to the data is improved sufficiently to justify the additional
complexity.  The situation is different when the goal is to 
produce a matched filter for a search for
a harmonically related frequency, as in the search for a signal at
half of the $\sim$580~Hz dominant brightness oscillation in
4U~1636--536 (Miller 1999).  For that purpose, it is not necessary to
justify the extra parameters used in the construction of the filter,
if no reference is made to the signal for which one is searching.
In the case of the search for the $\sim$290~Hz oscillation,
a five-parameter matched filter was used for each burst; matched
filters with fewer parameters also give a clear signal at
$\sim$290~Hz, although with lower significance because the filter
does not fit the data as well.

A general method to find the best-fit values of parameters and
their confidence regions employs a likelihood function.  In this
approach, we suppose that we have a model in which the countrate
as a function of time is predicted to be $s(t)$, from which we
can predict the number of counts $s_i$ in one particular bin $i$
of the data, which in this case is 1/8192~s in duration.  In 
general, $s_i$ is not an integer.  The actual number of counts
observed in bin $i$ is $c_i$, which is an integer.  With these
definitions, the Poisson likelihood of the full data set given
the model $s(t)$ is
\begin{equation}
{\cal L}=\Pi {s_i^{c_i}\over{c_i!}}e^{-s_i}\; ,
\end{equation}
where the product is over all of the bins of the data.  Note that
in normal applications of the point likelihood the bin sizes would
be so small that a given bin would have either zero or one
count, but the fixed bin size of 1/8192~s combined with the
high count rates during the bursts (up to $\sim$30,000~c/s; 
see Figure~\ref{figLightcurves}) means that many of the bins
have multiple counts.  The likelihood is maximized to determine
the best values of the parameters of the model waveform $s(t)$,
and approximate confidence contours can be estimated using
contours of constant log likelihood: $2\Delta\log{\cal L}=\Delta
\chi^2$ (Eadie et al.\ 1971, \S~9.4.3, p. 207).

The model waveform $s(t)$ will in general include components
related to the relatively slow change in the brightness of
the source as well as components related to the high-frequency
brightness oscillation.  However, the frequency scales are
different enough ($\sim$1--5~Hz for the slowly rising component
versus $\sim$580~Hz for the fast oscillations) that the fitting
of the two components are nearly independent of each other.  This
means that, when we analyze the behavior of the brightness
oscillations, we can simplify by assuming that the burst has
a constant average brightness.  With this in mind, the model
we consider is
\begin{equation}
s(t)=c_{\rm av}\left(1+A\cos 2\pi[\nu(t)t+\phi_0]\right)\; ,
\end{equation}
where $c_{\rm av}$ is the average countrate, $A$ is the
amplitude of the signal (which we assume to be time-independent),
$\nu(t)$ is the frequency as
a function of time, and $\phi_0$ is the phase of the
oscillation at the beginning of the data interval analyzed.

In this section we explore four different models for the
frequency behavior in the initial 0.75 seconds of each of
the four bursts (this time was chosen to conform to the
analysis of Miller [1999], which was performed to look for
a weaker $\sim$290~Hz oscillation).  The four models are:
\begin{equation}
\nu_1(t)=\nu_0
\end{equation}
\begin{equation}
\nu_2(t)=\nu_0+{\dot\nu}t
\end{equation}
\begin{equation}
\begin{array}{rl}
\nu_3(t) & =\nu_1+{\dot\nu_1}t\;,\;\; t<t_{\rm break}\\
      & =\nu_2+{\dot\nu_2}t\;,\;\; t>t_{\rm break}
\end{array}
\end{equation}
where continuity of the frequency is imposed, so that there
are four independent parameters, and finally
\begin{equation}
\begin{array}{rl}
\nu_4(t) & =\nu_1+{\dot\nu_1}t\;,\;\; t<t_{\rm break}\\
      & =\nu_2+{\dot\nu_2}t\;,\;\; t>t_{\rm break}
\end{array}
\end{equation}
where continuity of the frequency is not imposed, so that there
are five independent parameters.

For the purposes of this section the most interesting of
the parameters of the model waveform $s(t)$ is the frequency,
as opposed to the amplitude or the initial phase of the
brightness oscillation.  If the amplitude $A\ll 1$, as it
is in this case, then a tremendous speed-up in the search
procedure is possible with the use of the cross-correlation
(see, e.g., Helstrom 1960 or Wainstein \& Zubakov 1962 for
details of cross-correlation and matched filtering techniques)
\begin{equation}
H=C\left|\int_{t_0}^{t_0+T}c(t)e^{-i\nu(t)t}dt\right|^2\; ,
\end{equation}
where $t_0$ is the start time of the burst, $T$=0.75~s is
the duration of the burst, and $C$ is a normalization constant.
In practice this integral is actually calculated as a sum over
all of the bins of the data, and $dt$=1/8192~s is the duration
of a bin.  If $C=2/N_{\rm tot}$, where $N_{\rm tot}$ is the total
number of counts in the data set, then $H$ has the same
statistical properties as the Leahy power; $H$ is also related
to the $Z^2$ statistic used in pulsar period searches
(Buccheri et al.\ 1983; see Strohmayer \& Markwardt 1999 for a recent
use in the characterization of brightness oscillations during
thermonuclear X-ray bursts).  To lowest order in the oscillation
amplitude $A$ this description is mathematically identical to
the likelihood description, but it is much faster to apply because no
search need be performed for the amplitude or oscillation phase.
It is therefore preferable for low-amplitude oscillations. 

With this formalism, we can estimate the best values and
uncertainty regions for the different frequency models above.
The figures in the previous section, which were constructed using
a constant-frequency waveform, give this information 
for the one-parameter, constant-frequency
model.

\subsection{Two-Parameter Frequency Model}

The best-fit values for the two-parameter frequency model are
given in Table~2.  To estimate uncertainties on
these parameters, we performed a Monte Carlo analysis in which
we selected $10^6$ random values per burst of $\nu_1$,
${\dot\nu}_1$, ${\dot\nu}_2$, and $t_{\rm break}$, uniformly
sampled from, respectively, 576~Hz to 585~Hz; -12~Hz~s$^{-1}$
to 12~Hz~s$^{-1}$; -12~Hz~s$^{-1}$ to 12~Hz~s$^{-1}$; and 0~s
to 0.75~s.  The quoted uncertainties for single parameters
were computed using a Bayesian viewpoint, in which the posterior
probability density
was calculated throughout the interval and then integrated over
the other three parameters to produce a marginalized probability
distribution.  We have assumed a uniform prior probability
density over the whole space searched.  This means that the
posterior probability density is simply proportional to the
likelihood.  These confidence regions, which are the smallest
regions that encompass 68\% of the probability, are given in 
Table~3.  In some cases the maximum likelihood
value of a parameter obtained by extremization in the full two-dimensional
parameter space is outside the marginalized 68\% confidence
region.  This is symptomatic of the fact that the 
parameters are constrained only weakly by the data.

\begin{table}[t]
\centering
\begin{tabular}{ccc}
\multicolumn{3}{c}{\bfseries Table~2: Best-Fit Parameters}
{\rule[-2mm]{0mm}{6mm}}\\
\multicolumn{3}{c}{\bfseries for
Two-Parameter Model}{\rule[-2mm]{0mm}{6mm}}\\
\tableline
\tableline
Burst&$\nu_1$ (Hz)&${\dot\nu}_1$ (Hz s$^{-1}$)\\
\tableline
a&579.0& 4.0\\
b&580.3& 2.0\\
c&581.0&-3.8\\
d&578.6& 4.2\\
e&581.1&-3.4\\
\tableline

\end{tabular}
\end{table}

\begin{table}[ht]
\centering
\begin{tabular}{ccc}
\multicolumn{3}{c}{\bfseries Table~3: 68\% Confidence Regions}
{\rule[-2mm]{0mm}{6mm}}\\
\multicolumn{3}{c}{\bfseries for
Two-Parameter Model}{\rule[-2mm]{0mm}{6mm}}\\
\tableline
\tableline
Burst&$\nu_1$ (Hz)&${\dot\nu}_1$ (Hz s$^{-1}$)\\
\tableline
a&579.7--580.1&-8.0--8.0\\
b&580.9--581.2&-8.8--8.0\\
c&579.2--579.7&-8.0--8.0\\
d&578.8--579.9&-8.0--8.0\\
e&579.2--583.6&-8.0--8.0\\
\tableline

\end{tabular}
\end{table}

\subsection{Four-Parameter Frequency Model}

The best-fit values for the four-parameter frequency model are
given in Table~4.  As for the two-parameter
model, the uncertainties were estimated by marginalizing over
all but the parameter of interest; the confidence regions
containing 68\% of the probability are given in
Table~5.

\begin{table*}[t]
\centering
\begin{tabular}{ccccc}
\multicolumn{5}{c}{\bfseries Table~4: Best-Fit Parameters for
Four-Parameter Model}{\rule[-2mm]{0mm}{6mm}}\\
\tableline
\tableline
Burst&$\nu_1$ (Hz)&${\dot\nu}_1$ (Hz s$^{-1}$)
&${\dot\nu}_2$ (Hz s$^{-1}$)&$t_{\rm break}$ (s)\\
\tableline
a&581.3&-9.0&11.2&0.28\\
b&579.9& 4.0&0.5&0.32\\
c&579.1& 8.7&-6.7&0.20\\
d&579.7&-2.4&7.9&0.28\\
e&583.2&-12.0&-0.8&0.28\\
\tableline

\end{tabular}
\end{table*}

\begin{table*}[ht]
\centering
\begin{tabular}{ccccc}
\multicolumn{5}{c}{\bfseries Table~5: 68\% Confidence Regions for
Four-Parameter Model}{\rule[-2mm]{0mm}{6mm}}\\
\tableline
\tableline
Burst&$\nu_1$ (Hz)&${\dot\nu}_1$ (Hz s$^{-1}$)
&${\dot\nu}_2$ (Hz s$^{-1}$)&$t_{\rm break}$ (s)\\
\tableline
a&579.0--581.0&-8.8--2.4&2.4--6.4&0.16--0.42\\
b&578.8--580.6&-0.8--8.0&-4.0--3.2&0.13--0.63\\
c&579.0--581.6&-7.2--5.6&-5.6--4.0&0.13--0.64\\
d&578.0--579.8&-4.8--5.6&-4.0--6.4&0.11--0.64\\
e&579.2--582.0&-8.0--4.0&-5.6--4.8&0.14--0.63\\
\tableline

\end{tabular}
\end{table*}

\subsection{Five-Parameter Frequency Model}

The best-fit values for the five-parameter frequency model are
given in Table~6.  As for the two-parameter
model, the uncertainties were estimated by marginalizing over
all but the parameter of interest; the confidence regions
containing 68\% of the probability are given in
Table~7.

\begin{table*}[t]
\centering
\begin{tabular}{cccccc}
\multicolumn{6}{c}{\bfseries Table~6: Best-Fit Parameters for
Five-Parameter Model}{\rule[-2mm]{0mm}{6mm}}\\
\tableline
\tableline
Burst&$\nu_1$ (Hz)&$\nu_2$ (Hz)&${\dot\nu}_1$ (Hz s$^{-1}$)
&${\dot\nu}_2$ (Hz s$^{-1}$)&t$_{\rm break}$ (s)\\
\tableline
a&581.2&578.0&-7.6&6.4&0.34\\
b&579.0&579.6& 8.0&3.2&0.34\\
c&578.6&578.0& 6.8&0.8&0.41\\
d&578.0&582.6&-0.4&-5.2&0.25\\
e&582.0&585.0&-8.0&-0.4&0.41\\
\tableline

\end{tabular}
\end{table*}

\begin{table*}[ht]
\centering
\begin{tabular}{cccccc}
\multicolumn{6}{c}{\bfseries Table~7: 68\% Confidence Regions for
Five-Parameter Model}{\rule[-2mm]{0mm}{6mm}}\\
\tableline
\tableline
Burst&$\nu_1$ (Hz)&$\nu_2$ (Hz)&${\dot\nu}_1$ (Hz s$^{-1}$)
&${\dot\nu}_2$ (Hz s$^{-1}$)&t$_{\rm break}$ (s)\\
\tableline
a&579.6--581.8&577.4--581.6&-8.8--4.8&0.0--6.4&0.11--0.48\\
b&577.8--580.6&579.4--582.0&-5.6--8.8&-2.4--3.2&0.08--0.52\\
c&577.8--581.2&577.4--581.8&-6.4--8.8&-5.6--1.6&0.09--0.63\\
d&579.2--581.6&577.0--581.6&-9.6--6.4&-2.4--6.4&0.08--0.34\\
e&579.0--582.8&577.6--582.4&-8.8--6.4&-4.8--3.2&0.08--0.56\\
\tableline

\end{tabular}
\end{table*}

\subsection{Summary of Frequency Models}

\begin{table*}[t]
\centering
\begin{tabular}{ccccc}
\multicolumn{5}{c}{\bfseries Table~8: Relative Log Likelihoods for
Different Models}{\rule[-2mm]{0mm}{6mm}}\\
\tableline
\tableline
Burst&1-Param&2-Param&4-Param&5-Param\\
\tableline
a&0.0&1.3&4.3&4.5\\
b&0.0&1.2&1.6&2.8\\
c&0.0&1.3&2.0&3.1\\
d&0.0&2.2&3.2&5.4\\
e&0.0&0.6&1.0&2.0\\
\tableline

\end{tabular}
\end{table*}

The best-fit parameters and relative log likelihoods are
listed in Table~8; as indicated above,
$2\Delta\log{\cal L}\approx\Delta\chi^2$.
From this table, it is clear that for all but burst four
it is not necessary to use the five-parameter fit, and
for bursts 2, 3, and 4 it is not necessary to use a
model more complicated than the two-parameter model
in which the frequency and frequency derivative are
constant throughout the first 0.75 seconds.  For burst~(d)
by itself the five-parameter model is preferred
at only the $2\sigma$ level compared to the four-parameter
model, and for all five bursts combined the five-parameter
model is preferred at less than the $1\sigma$ level relative
to the four-parameter model.
For all five bursts combined, the four-parameter model
is preferred at less than the $1\sigma$ level compared
to the two-parameter model, and the two-parameter model
is preferred at less than the $2\sigma$ level compared
to the one-parameter model.

\section{DISCUSSION AND SUMMARY}

What can be learned from this detailed characterization of
the burst brightness oscillations in 
4U~1636--536?  The clearest impression left
is that there are no simple statements about the frequency
behavior that are true for all of the bursts.  In two of the
bursts, one can make an argument that the oscillation
frequency is initially 1--2~Hz below
the asymptotic frequency, and then rises.  In this interpretation,
the asymptotic frequency is extremely close to the spin
frequency of the neutron star.  This picture can be qualitatively
explained by the idea that the burning layer lifts
20--50 meters during the burst and settles down gradually.
However, the burst on 31 December 1996 does not follow
this pattern.  The frequency in the initial second is
indeed lower than the maximum value attained, but the
significance of this initial signal is low (Leahy power of 10).
The maximum is followed by a clear decrease in the frequency over
several seconds, with a total phase change equivalent to
more than five complete circuits around the star.  This
happens during a time when the countrate
decreases from approximately 2/3 of the maximum to approximately
1/3 of the maximum.  The burst on 29 December 1996 has a
very strong and significant brightness oscillation in
its tail, which appears to level out to a constant frequency.
However, near the peak of the light curve for this burst
the oscillation frequency is higher than this asymptotic
frequency, at a 3$\sigma$ significance level.

Such a drop in frequency is not expected in the simplest
version of the hypothesis that the frequency changes are
caused by the rise of the burning layers.  In this model,
the highest frequency should be observed when the layers
are fully coupled to the core of the star, which is expected
to occur when the frequency has reached its asymptotic
limit.

Another constraint on the hypothesis that the asymptotic
frequency equals the spin frequency (after correcting
for orbital Doppler
shifts) is that the variation in the observed asymptotic
frequency must be consistent with the possible modulation
due to the binary motion of the neutron star.  From binary
evolution theory (see, e.g., Lamb \& Melia 1987;
Verbunt \& van den Heuvel 1995), an LMXB such as 4U~1636--536 
with a 3.8~hr orbital
period (van Paradijs et al.\ 1990) that contains a neutron
star of mass $M_{\rm NS}$=1.4$M_\odot$ to 2.0$M_\odot$ has 
a companion star of mass $M_c\approx$0.4$M_\odot$.  Assuming that
the orbit is approximately circular, the orbital velocity
of the neutron star is therefore 90--130~km~s$^{-1}$, 
implying a maximum frequency modulation of $\Delta\nu/\nu
=4.3\times 10^{-4}$, or approximately 0.25~Hz if $\nu$=580~Hz.
Therefore, the observed asymptotic frequency cannot be different
by more than 0.5~Hz for two different bursts.  The analysis
of the 31 December 1996 burst reported in \S~2 indicates
that eight seconds after the start of the burst the frequency
is less than 579.0~Hz.  The asymptotic frequency in the
burst on 29 December 1996 is 581.43~Hz, so the frequency in
the 31 December 1996 burst must rise by 2~Hz to reach a
plausible spin frequency.

It is difficult to reconcile this frequency behavior with
what is expected in the simplest version of the rising burning layer
hypothesis.  One possibility is that the observed
frequency changes are not simply indicative of the spin
frequency of the burning layer, but also include a time-dependent
change in the phase at which the photons emerge relative
to the phase of the burning layer.  This would be observationally
indistinguishable from a pure frequency change, and would add
an extra degree of freedom to the model.

Even this, however, is subject to significant observational
restrictions.  To see this, consider the following observational
trends, which have been observed in many bursts from several
sources (see, e.g., Strohmayer et al.\ 1998 for a summary).  
In the remainder of this section we assume that all
quantities (e.g., frequencies, times, and phases) are
measured at infinity.

(1) There are several bursts in which burst
oscillations are seen for the entire burst, and
do not disappear during the time of peak countrate.

(2) Aside from an early phase in which there may
be a frequency decrease, the frequency increases
smoothly as the burst progresses.

(3) The total phase lag of the oscillations
compared with a hypothetical oscillation that
has a constant frequency equal to the frequency
in the burst tail is as much as $10\pi$.

The total amount of energy in a burst is $\sim 10^{39}$~ergs.
If expansion of a layer and angular momentum conservation
are to explain the $\sim$0.3\%--1\% change in the observed
angular frequency, then the layer must rise by a distance
that is a fraction $\sim$0.2\%--0.5\% of the radius of
the star, or 20 to 50 meters.  The surface gravity of
a neutron star is $\sim 2\times 10^{14}$~cm~s$^{-2}$, so
the largest amount of mass that can be lifted to the
required 20--50 meter height above the surface is
$\sim$1--2$\times 10^{21}$~g.  If most of the $\sim 10^{13}$~cm$^2$
surface area of the star is involved, this implies that the
greatest column depth which could be lifted to the required
height is roughly $10^8$~g~cm$^{-2}$, which is comparable
to the expected $10^6-10^8$~g~cm$^{-2}$depth of ignition 
(see, e.g., Fushiki \& Lamb 1987; Brown \& Bildsten 1998).

One may therefore distinguish two
scenarios: (1)~the burning layer rotates with the core
of the star at a constant spin frequency and the observed
frequency shifts are caused by phase shifts induced by
radiation transport through more slowly rotating layers,
and (2)~the burning layer itself is lifted and rotates more
slowly than the core of the star.  We now treat these in
order. 

Suppose for simplicity that the burning layer has an infinitesimal
vertical extent, that it has some restricted azimuthal extent,
and that it all rotates with the same angular frequency 
$\omega_{\rm burn}(t)$.
The energy from this layer propagates upwards through the
atmosphere, which in general may be composed of layers with
different angular frequencies.  Therefore, the phase of
emergence of the radiation may differ from the phase of the
burning layer at the time of the emission of the radiation.
Under the rising burning layer hypothesis, it is expected that
the angular frequency of higher layers is less than the
angular frequency of lower layers ($d\omega/dh<0$).  Hence, there is expected
to be a lag $\phi_{\rm lag}>0$ between the phase of emergence
and the phase of emission.  This phase lag will, in general,
have a time-dependence, as the scale height of the atmosphere
and the angular frequency of different layers in the atmosphere
changes throughout the burst.  An observer at infinity will
therefore see a net angular frequency of a hot spot that is
equal to $\omega_{\rm burn}(t)-{\dot\phi}_{\rm lag}(t)$.

Consider first a burning layer that rotates with the stellar
core throughout the burst.  Then $\omega_{\rm burn}(t)$
=const=$\omega_{\rm spin}$.  If neither
$\omega(h)$ nor the density or height of the envelope
changes with time, then $\phi_{\rm lag}$ is a constant and
the observed frequency is just $\omega_{\rm spin}$.
Hence, in order to have an apparent frequency shift in
this situation, the structure or angular velocity of the
envelope must change with time.

Now consider an envelope that does change with time.
For us to observe a frequency less than $\omega_{\rm spin}$,
it is necessary that ${\dot\phi}_{\rm lag}(t)>0$, so
the characteristic phase of emergence of the radiation
must lag the phase of the source of heat by a greater and
greater amount with increasing time (the increase of this phase lag
with time must itself decrease with time to produce the observed
increase in frequency).  But how is this possible?  As the
envelope settles down, the phase lag should {\it decrease},
because $d\omega(t)/dh<0$.
But if the phase lag decreases, the observed
frequency should be {\it higher} than the spin frequency.  This
is not seen in most bursts, and even in the burst on 29 December 1996
where there does appear to be a short period of spindown, the total
phase lead implied by the spindown is much smaller than the total
phase lag implied by the spinup near the beginning of the burst.
Thus, the preceding set of assumptions is inconsistent with the data.

This demonstrates that the observed frequency behavior is
inconsistent with the source of heat (i.e., the burning
layer) rotating at
a constant frequency equal to $\omega_{\rm spin}$.  Instead,
the source of heat must change its frequency during the burst.

To analyze this situation, let us now consider a burning
layer with a finite thickness, so that the observed photons
are a superposition of the photons from many infinitesimal
layers such as discussed above.  The observed frequency of
oscillation is then a superposition of the frequencies due
to the infinitesimal layers.

Consider two of these infinitesimal slices, labeled 1 and 2,
where slice 1 is higher than slice 2.  Suppose that these
slices are not coupled to each other.  Then, by assumption, the
angular frequency $\omega_{\rm burn,1}$ of slice 1 is less than
the angular frequency $\omega_{\rm burn,2}$ of slice 2.  
In addition, because the
photons from slice 2 have to travel through the same atmospheric
layers as the photons from slice 1 in addition to the layers
between 2 and 1, the phase lag $\phi_{\rm lag,1}$ of photons
from slice 1 is expected to be less than the phase lag 
$\phi_{\rm lag,2}$ of photons from slice 2.  Hence, as the 
atmospheric scale height decreases, it is expected that
$\phi_{\rm lag,2}$ will decrease more rapidly than
$\phi_{\rm lag,1}$ does, so that
\begin{equation}
{\dot\phi}_{\rm lag,2}<{\dot\phi}_{\rm lag,1}<0\; .
\end{equation}
Therefore, the difference between the angular frequency of
the photons from slice 2 and the angular frequency of the
photons from slice 1 is
\begin{equation}
\omega_{\rm burn,2}-\omega_{\rm burn,1}+
{\dot\phi}_{\rm lag,1}-{\dot\phi}_{\rm lag,2}>
\omega_{\rm burn,2}-\omega_{\rm burn,1}\; .
\end{equation}
This means that the phases
of emergence of radiation diverge rapidly from each other, 
which leads quickly
to a low amplitude unless the heat source has a small vertical
extent.  The requirement that the amplitude be significant
means that the total azimuthal phase subtended by the emergent
radiation has to be much less than $2\pi$.  The integrated
phase lag relative to the stellar core is often $10\pi$ or
larger, hence the average vertical extent of the heat source
must be much less than 1/5 of the vertical distance from the
original location of the heat source to its location during
the burst.  An alternative to having the vertical extent of
the layer be small is that the burning layer may be tightly
coupled to itself, so that its angular frequency is 
approximately constant over a significant vertical distance.

To summarize, several conclusions may be drawn about the standard model for
frequency changes during burst oscillations, which we take to
be the picture that at least part of the
burning layer is lifted and then settles gradually to the surface as
the flux drops, producing an observed asymptotic frequency equal to the
spin frequency of the neutron star Doppler-shifted by the
orbital motion of the neutron star.  (1)~The burning region
itself (and not just overlaying optically thick layers) must
be lifted by 20--50 meters from the surface, (2)~this region
must remain decoupled from the rest of the star, presumed to
be rotating at the original spin frequency, for several seconds,
(3)~to produce the observed coherence of the brightness oscillations
during the rise in frequency, the burning layer must either have
a vertical extent much smaller than its height above the surface
or be strongly coupled to itself to prevent relative azimuthal
motion, and (4)~the existence of a frequency greater than the
asymptotic frequency (as in the 29 December 1996 burst) implies
that something other than differential rotation (e.g., variation
in the phase lag) must account for
at least part of the observed frequency change.  The prolonged
decrease in frequency in the tail of the 31 December 1996 burst
is not straightforwardly fit into this picture.

Despite these difficulties, the high stability (Strohmayer et al.\ 1998)
and coherence (Markwardt \& Strohmayer 1999) of the brightness
oscillations in the tails of bursts from sources such as 
4U~1728--34 argue persuasively that the frequency in the tail
of the bursts is close to either the fundamental or the first overtone
of the neutron star spin frequency.  Moreover, the general
picture in which frequency changes are attributed to changes in the height
of the emitting layer accounts approximately for the magnitude of the
frequency change and explains why the frequency tends to
rise near the beginning of the burst.  However, in its current
form it suffers from apparently serious problems.  It is extremely
important that there be a detailed investigation of, e.g., the coupling
between differentially rotating layers, and that other ideas
be explored so that the strengths and weaknesses of the rising
layer model are put into sharper focus.

\acknowledgements
We thank Don Lamb and Fred Lamb for discussions about models
of the frequency change, and Don Lamb, Dimitrios Psaltis, and
Carlo Graziani for comments on a previous version of this paper.  
This research has made use of data
obtained through the High Energy Astrophysics Science Archive
Research Center Online Service, provided by the NASA/Goddard
Space Flight Center.  This work was supported in part by NASA grant
NAG~5-2868, NASA AXAF contract SV~464006, and NASA ATP grant
number NRA-98-03-ATP-028.


\begin{references}

\reference{B95} Bildsten, L. 1995, ApJ, 438, 852

\reference{B98} --------- 1998, in ``The Many Faces of
Neutron Stars", ed. R.~Buccheri, A.~Alpar, \& J. van Paradijs
(Dordrecht: Kluwer), p. 419

\reference{BB98} Brown, E.~F., \& Bildsten, L. 1998, ApJ, 496, 915

\reference{B83} Buccheri, R. 1983, A\& A, 128, 245

\reference{E71} Eadie, W.~T., Drijard, D., James, F.~E.,
Roos, M., \& Sadoulet, B. 1971, Statistical Methods in
Experimental Physics (Amsterdam: North-Holland)

\reference{FL99} Fox, D.~W., \& Lewin, W.~H.~G. 1999, IAUC 7081

\reference{FW92} Fryxell, B.~A., \& Woosley, S.~E. 1982,
ApJ, 258, 773

\reference{FL87} Fushiki, I., \& Lamb, D.~Q. 1987, ApJ, 323, L55

\reference{H60} Helstrom, C.~W. 1960, Statistical Theory
of Signal Detection (New York: Pergamon)

\reference{J78} Joss, P.~C. 1978, ApJ, 225, L123

\reference{LM87} Lamb, D.~Q., \& Melia, F. 1987a, ApJ, 321, L133

\reference{L83} Leahy, D.~A., Darbro, W., Elsner, R.~F.,
Weisskopf, M.~C., Sutherland, P.~G., Kahn, S., \&
Grindlay, J.~E. 1983, ApJ, 266, 160

\reference{LB82} Livio, M., \& Bath, G.~T. 1982, A\& A, 116, 286

\reference{M99} Miller, M.~C. 1999, ApJ, 515, L77

\reference{N84} Nozakura, T., Ikeuchi, S., \& Fujimoto, M.~Y.
1984, ApJ, 286, 221

\reference{R81} Ruderman, M. 1981, Prog. Part. Nucl. Phys., 6, 215

\reference{S82} Shara, M.~M. 1982, ApJ, 261, 649

\reference{SM99} Strohmayer, T.~E., \& Markwardt, C.~B. 1999,
ApJ Lett., accepted (astro-ph/9903062)

\reference{S98a} Strohmayer, T.~E., Swank, J.~H., \&
Zhang, W. 1998a, in Proceedings of the Symposium ``The
Active X-Ray Sky: Results from BeppoSAX and Rossi-XTE", Rome, Italy,
21-24 October, 1997, Nuclear Physics B Proceedings Supplements. Eds.
L.~Scarsi, H.~Bradt, P.~Giommi, and F.~Fiore

\reference{SZS97} Strohmayer, T.~E., Zhang, W., \& Swank, J.~H.
1997, ApJ, 487, L77

\reference{S98b} Strohmayer, T.~E., Zhang, W., Swank, J.~H.,
\& Lapidus, I. 1998b, ApJ, 503, L147

\reference{S98c} Strohmayer, T.~E., Zhang, W., Swank, J.~H.,
White, N.~E., \& Lapidus, I. 1998c, ApJ, 493, L135

\reference{vP90} van Paradijs, J. et al.\ 1990, A\&A,
234, 181

\reference{VH95} Verbunt, F., \& van den Heuvel, E.~P.~J. 1995, in
X-Ray Binaries, ed. W.~H.~G. Lewin, J. van Paradijs, \&
E.~P.~J. van den Heuvel (Cambridge U. Press, Cambridge), 457

\reference{WZ62} Wainstein, L.~A., \& Zubakov, V.~D. 1962,
Extraction of Signals from Noise (Englewood Cliffs: Prentice-Hall)

\end{references}
\end{document}